\documentclass[12pt]{iopart}
\usepackage{graphicx}
\usepackage{dcolumn}
\usepackage{iopams}
\usepackage{amssymb}
\usepackage{color}
\usepackage{subfig}

\expandafter\let\csname equation*\endcsname\relax 
\expandafter\let\csname endequation*\endcsname\relax 
\usepackage{amsmath}

\newcommand{\bvec}[1]{\ensuremath{\mathbf{#1}}}

\begin{document}

\title[Importance of $d$-$p$ Coulomb interaction]
{Importance of $d$-$p$ Coulomb interaction for high T$_C$ cuprates and other oxides}

\author{P.~Hansmann$^1$, N.~Parragh$^2$, A.~Toschi$^3$, G.~Sangiovanni$^2$, and K.~Held$^3$}
\address{$^1$Centre de Physique Th{\'e}orique,
Ecole Polytechnique, CNRS-UMR7644, 91128 Palaiseau, France\\
$^2$ Universit\"at W\"urzburg, Am Hubland, D-97074 W\"urzburg, Germany\\
$^3$Institut f\"ur Festk\"orperphysik, Vienna University of Technology, 1040 Vienna, Austria
}


\begin{abstract}
Current theoretical studies of electronic correlations in
transition metal oxides typically only account for the local
repulsion between  $d$-electrons even if oxygen ligand $p$-states are an
explicit part of the effective Hamiltonian. Interatomic interactions such as
$U_{pd}$ between $d$- and (ligand) $p$-electrons, as well as the local interaction between $p$-electrons, are neglected. Often, the relative $d$-$p$ orbital splitting has to be adjusted ``ad hoc'' on the basis of the experimental evidence.
By applying the merger of local density approximation and dynamical mean field
theory (LDA+DMFT) to the prototypical case of the $3$-band Emery $dp$ model for the cuprates, we demonstrate that, without any ``ad hoc'' adjustment of the orbital splitting,  the charge transfer insulating state is stabilized
by the interatomic interaction $U_{pd}$. 
 Our study hence shows how to improve  realistic material calculations that explicitly include the $p$-orbitals. 

\end{abstract}

\maketitle

\section{Introduction} 
A primary effect of electronic correlations is to turn materials,
which according to band-theory would be metallic, into insulators
classified as Mott-Hubbard and charge transfer
insulators\cite{ZSA}.
 Experimentally, various transition metal oxides were intensively investigated\cite{imada98}. The huge interest in this subject goes back to the unexpected discovery\cite{Bednorz86} of high temperature superconductivity,  by doping the charge-transfer insulating cuprates. On the theoretical side, great efforts to solve the Hubbard model\cite{Hubbardmodel} have been made since this model captures, in a fundamental way, the competition between electronic mobility and localization, generated by a purely local interaction. 

No exact solution of the Hubbard model in two or three dimension is  known hitherto,
but a  breakthrough was 
achieved 
by dynamical mean field theory (DMFT) \cite{Metzner89a,Georges92a,Georges96a}. DMFT extends the concept of a mean field theory to quantum mechanics and takes into account, non perturbatively, a major part of the electronic correlations, i.e., the local electronic correlations in time. 
A further, and significant, advantage of DMFT is the possibility to combine it
with {\sl ab-initio} density functional calculations, e.g., with the local density approximation (LDA) in the so-called LDA+DMFT\cite{DFTDMFT} approach, which has already evolved to a considerable maturity. By means of LDA+DMFT, realistic calculations of correlated materials with a high degree of accuracy became possible, as it is well exemplified by the successful treatment of the Mott-Hubbard transition in V$_2$O$_3$ \cite{noteV2O3}.

The high-temperature superconducting cuprates obviously represent a much more
challenging case for DMFT-based methods. In fact, the extremely anisotropic
(quasi two-dimensional) layered structure of these compounds enhances the
importance of {\sl non-local} spatial correlations. Hence, if aiming at the
low-temperature physics of cuprates, non-local correlations
need to be taken into account, which could be achieved by
cluster\cite{CDMFT} or diagrammatic extensions\cite{diag_ext} of DMFT.
Here, we are not focusing on if, or up to what extent, the Hubbard model describes the unconventional superconductivity of the cuprates -- a question on which no consensus has been reached yet. 
Instead, we will focus on the analysis of the high-temperature regime of the
undoped cuprates. The effect of non-local
correlations will be gradually mitigated by temperature\cite{Gull2D,DGAresul,Gull3D},  employing LDA+DMFT is more justified. LDA+DMFT was already successfully applied to the cuprates for analyzing the anomalies  in the restricted optical sum rules\cite{sumoptexp,Toschi,Comanac,Nicoletti} and the fingerprints of the Zhang-Rice physics in photoemission spectra\cite{LucaWang,Weber2008}. 
Here, we will employ LDA+DMFT \footnote{Specifically, for all the DMFT and DMFT+Hartree calculations presented in this work the W\"urzburg-Wien ``w2dynamics'' code\cite{Parragh2012}, exploiting the hybridization-expansion implementation\cite{Werner2006}  of the continuous-time quantum Monte Carlo method\cite{Rubtsov2004}, has been used.} to investigate the fundamental nature of the high-temperature insulating state in the undoped compounds. This insulating state, classified as a charge transfer insulator\cite{imada98}, is in fact common for many transition metal oxides.
A crucial step in the LDA+DMFT method is the derivation of the low energy Hamiltonian from ab initio calculations by renormalization (downfolding \cite{downfolding}) to certain basis states around the Fermi energy. 
The decision on the basis, in fact, is crucial. For the cuprates this question bascially boils down to include oxygen $2p$ ligand-states explicitly or implicitly. The simplest possible Hamiltonian with both Cu $3d$ and oxygen $2p$ states is the Emery model which includes, besides the Cu $x^2-y^2$ band, the strongest hybridizing $p_x$ and $p_y$ states in the $xy$ plane (an extensive discussion on the downfolding procedure for cuprates can be found in Ref.\cite{andersen95}).


In the renormalization sense, physical observables of different effective Hamiltonians should coincide at low energies, albeit the larger effective Hamiltonians might be more accurate than those defined on narrower energy windows. By extending the basis set, one  (i) enlarges the energy range and the physical processes explicitly included in the downfolding procedure, and (ii) obtains a better localization of the more correlated $d$-orbitals.   
Such a general statement is however premature if the extension exclusively relies on the one-particle (``kinetic'') part of the Hamiltonian. One should not forget that the Coulomb interaction needs to be renormalized as well. 

The material best studied in LDA+DMFT within different basis-sets  is arguably LaNiO$_3$,  and  heterostructures thereof. Here, including the $p$ states in a  $dp$-model\cite{MillisLaNiO3}  leads to qualitatively different results compared to the $d$-only model\cite{HansmannLaNiO3}.
As pointed out in Ref.\cite{explanationLaNiO3}, the physics is very different in both cases: in the $dp$-model the $d$-occupancy is larger with a  tendency of two $d$-electrons  to form a spin-1 due to Hund's exchange. On the contrary, the $d$-only model is built with a (fixed) $d^1$ configuration. In such a situation, it is pretty unclear, whether the $dp$-model does actually provide a more accurate description of the physics, as one might naively expect due to the larger number of degrees of freedom explicitly taken into account within the larger basis-set of the downfolded model \cite{rueeggNi2013}.

 Hitherto,  the additional oxygen $p$-orbitals in the extended basis-sets have been considered as {\sl not} interacting in state-of-the-art LDA+DMFT calculations, i.e., only a local $d$-$d$ interaction has been taken into account. 
While the oxygen $p$-orbitals are, due to their nodeless radial wave function also quite localized, their typical filling in transition metal oxides renders the onsite $p$-interaction to be non-crucial.
Yet, at the same time, our numerical results  show that the effect of the \textit{interatomic} $2p$-$3d$ interaction, in the following coined as $U_{pd}$, {\sl does} have a large effect. $U_{pd}$ is an essential parameter that  controlls the metal-to-insulator transition, as well as the size of the charge transfer gap. 

The paper is organized as follows: In Section \ref{Sec:Emery} the construction of the Emery model is described starting from the full LDA bandstructures. Section  \ref{Sec:methods} and the Appendix are dedicated to the  DMFT and Hartree treatment of the various Coulomb interactions.  Section  \ref{Sec:results} presents the results obtained, in particular the opening of the charge transfer gap and the importance of $U_{pd}$ to this end. Finally,  Section \ref{Sec:conclusion} summarizes the paper.

\section{The Emery model revisited} 
\label{Sec:Emery} 
A popular tight--binding Hamiltonian for the high T$_{\rm C}$ cuprates, including oxygen $p$-states, is a 3--band model suggested by Emery\cite{emery87} in 1987. The Emery model consists of one planar Cu $x^2-y^2$ band, two oxygen $p_x$ and $p_y$ bands and takes into account a $d$--$p$ hopping, see Table\  \ref{emeryham}. It is thus the minimal model in order to describe the charge transfer insulating state and, moreover, the physics of a Zhang-Rice singlet \cite{zhang_rice}. These features are obviously beyond a description of the cuprates within an effective single band model.
However, Andersen \emph{et al.} \cite{andersen95} concluded from downfolding the \emph{ab initio} LDA bandstructure that the original model as described in Ref.~\cite{emery87} should be extended. These extensions  originate basically from the inclusion of an \emph{axial} degree of freedom (including mostly Cu 4s states and apical oxygen states) which was seen to describe the material--dependence of the electronic structure in the cuprates\cite{pavarini01}. The derivation starts from eight bands which are separated into a $4 \times 4$ $\sigma$--bonding block of Cu $3d_{x^2-y^2}$, O$_{1}$ 2p$_x$, O$_{2}$ $2p_y$, as well as Cu $4s$ (with some Cu $3d_{3z^2-r^2}$ character), and another $4\times 4$ $\pi$--bonding block of Cu-$3d_{xz}$, Cu $3d_{yz}$, O$_{1}$ $2p_z$, and O$_{2}$ $2p_z$. The generic Hamiltonian for the CuO$_2$ planes is the $4\times 4$ $\sigma$--block, since it contains the conduction band. Due to symmetry reasons, the $\sigma$-- and the $\pi$--block do not hybridize in the limit of \emph{flat planes} \cite{andersen95}.

\begin{table*}\renewcommand\arraystretch{1.7}
  \begin{equation*}
    \hat{H}^{\rm NMTO}=\sum_{\bvec{k}lm\sigma}h^{\rm LDA}_{lm}(\bvec{k})c^{\dagger}_{l\sigma}(\bvec{k})c_{m\sigma}(\bvec{k})
  \end{equation*}
  
  \begin{equation*}
    h^{\rm NMTO}(\bvec{k})\left(\begin{array}{ccc}
        h_d(\bvec{k})& h_{d,px}(\bvec{k})& h_{d,py}(\bvec{k})\\
        h_{px,d}(\bvec{k})& h_{px}(\bvec{k})& h_{px,py}(\bvec{k})\\
        h_{py,d}(\bvec{k})& h_{py,px}(\bvec{k})& h_{py}(\bvec{k})
        \end{array}\right)
  \end{equation*} 

  \begin{equation*}
    \begin{array}{lcl}
      h_{d}(\bvec{k})&=&\varepsilon_d+2{\rm t}_{dd}({\rm Cos}(k_x)+{\rm Cos}(k_y))+4{\rm t}'_{dd}{\rm Cos}(k
_x){\rm Cos}(k_y)\\
      h_{px}(\bvec{k})&=&\varepsilon_p+2.0({\rm t}'_{pp}{\rm Cos}(k_x)+{\rm t}''_{pp}{\rm Cos}(k_y)+2{\rm t}'''_{pp}{\rm Cos}(k_x){\rm Cos}(k_y))\\
      h_{py}(\bvec{k})&=&\varepsilon_p+2.0({\rm t}'_{pp}{\rm Cos}(k_y)+{\rm t}''_{pp}{\rm Cos}(k_x)+2{\rm t}'''_{pp}{\rm Cos}(k_y){\rm Cos}(k_x))\\[0.2cm]
      h_{d,px}(\bvec{k})&=&2(({\rm t}_{pd}+2{\rm t}'_{pd}{\rm Cos}(k
_y)){\rm Sin}(k_x/2)+({\rm t}''_{pd}+2{\rm t}'''_{pd}{\rm Cos}(k_y)){\rm Sin}(3k_x/2))\\
      h_{d,py}(\bvec{k})&=&-2.0(({\rm t}_{pd}+2{\rm t}'_{pd}{\rm Cos}(k_x)){\rm Sin}(k_y/2)+({\rm t}''_{pd}+2{\rm t}'''_{pd}{\rm Cos}(k_x)){\rm Sin}(3k_y/2))\\
      h_{px,py}(\bvec{k})&=&-4.0({\rm t}_{pp}{\rm Sin}(k_x/2){\rm Sin}(k_y/2)+{\rm t}''''_{pp}({\rm Sin}(3k_x/2){\rm Sin}(k_y/2)\\
      & & \;\;\;\;\;\;\;\;\;\;\;\;\;\;\;\;\;\;\;\;\;\;\;\;\;\;\;\;\;\;\;\;\;\;\;\;\;\;\;\;\;\;\;\;\;\;\;\;\;\;\;\;\;\;\;\;\;\;\;\;\;\;\; +{\rm Sin}(3k_y/2){\rm Sin}(k_x/2)))\\
    \end{array}
  \end{equation*}  
  \caption{Extended Emery model \cite{kent08} including $p$--$p$ hopping
    mediated by the \emph{material dependent} axial degree of
    freedom.}
  \label{emeryham}
\end{table*}

\begin{table*}
\begin{tabular*}{\columnwidth}{@{}r*{20}{@{\extracolsep{1pt plus22pt}}r}}

NMTO& $\varepsilon_d-\varepsilon_p$ & $t_{dd}$ & $t_{pd}$ & $t_{pd}'$ &$t_{pp}$ & $t_{pp}'$ & $t_{pp}''$ & $t_{pp}'''$\\
\hline\\[-0.2cm]
N=0& $0.43$ & $-0.10$ & $0.96$ & $-0.10$ &$0.15$ & $-0.24$ & $0.02$ & $0.11$\\
N=1& $0.95$ & $0.15$ & $1.48$ & $0.08$ &$0.91$ & $0.03$ & $0.15$ & $0.03$\\

\end{tabular*}
\caption{\label{tvaluestab} Hopping integrals for the Emery model Table \ref{emeryham} as obtained by 0MTO and 1MTO in \cite{Nuss,kent08}.}
\end{table*}

Starting from the four band $\sigma$--Hamiltonian one can
arrive either at i) a two-band Hamiltonian with \emph{planar} (''dressed'' Cu
$x^2-y^2$) and \emph{axial} (''dressed'' Cu $4s$) degrees of freedom \footnote{In this case the $\sigma$--oxygen states fold into the \emph{planar} $x^2-y^2$ conduction band.}, or at ii) the three band Emery model. The latter is obtained by folding the \emph{axial}  degrees of freedom down to the oxygen bands, resulting in an additional (material dependent) $p$--$p$ hopping  $t_{pp}$ and renormalization of the onsite $p$--energy. 

Quantitative Hamiltonians can be obtained by means of N$^{\rm th}$ order Muffin
Tin Orbitals (NMTO) \cite{downfolding} based on a L\"owdin downfolding scheme
with subsequent ''N-ization'' ($N=0$: linearization). This technique has been
already established and used in several LDA+DMFT calculations
\cite{LDADMFTcalc}. In our paper, we will consider two different low-energy
Hamiltonians \cite{Nuss,kent08}: The first one consists of $N=0$ muffin tin orbitals which have
been linearized around the Fermi energy. The 0MTO basis gives correct LDA wave
functions at the Fermi level and, by virtue of the variational principle, the
correct LDA Fermi surface and Fermi velocities. The 1MTO basis gives, in
addition to this, the correct LDA wave functions at an energy $E1$ chosen at the
bottom of the $pd$-bonding band, and hence, the correct $E(k)=E1$ surface and
velocities. The resulting Hamiltonian can be written analytically 
as in  Table~\ref{emeryham} with hopping parameters given in  Table~\ref{tvaluestab}.

\begin{figure*}%
\includegraphics[width=1.0\textwidth]{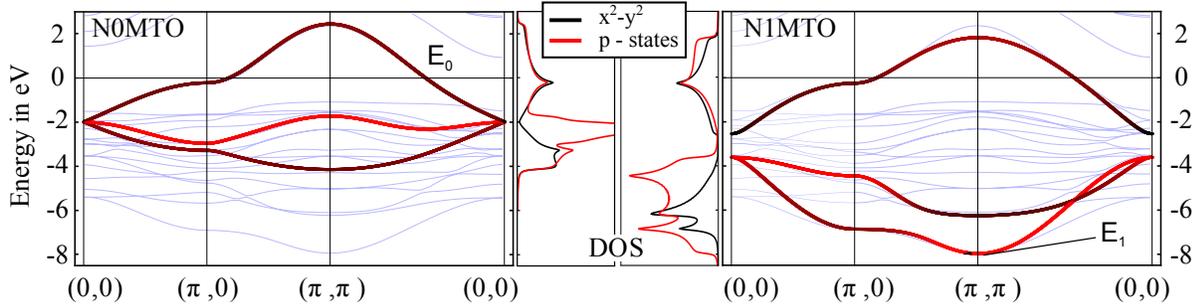}%

\caption{(color online) Bandstructure and single particle density of states for the  0MTO and 1MTO downfolded model\cite{Nuss,kent08}. The color codes the orbital character: 3d $x^2-y^2$ is black, whereas the $p$--states are summarized and plotted in red (grey) -- color mixing corresponds to a mixed orbital state. The LDA bands are shown as thin (blue) lines.}%
\label{NMTO_bands}%
\end{figure*}

In Fig.~\ref{NMTO_bands} we show these NMTO downfolded  bands for La$_2$CuO$_4$ from Kent
\emph{et al.} \cite{kent08} for $N=0$ (linearization around the Fermi energy - left hand side) and $N=1$ (with $E_1$ chosen to be at the bottom of the $pd$-bonding band at $(\pi,\pi)$ - right hand side). The corresponding hopping parameters can be found in Table\ \ref{tvaluestab}. The difference of the N=0 and N=1 models is the energy range in which they reproduce the cuprate LDA bandstructure (light blue lines): While the $N=0$ model only has one fixed energy, namely the
Fermi energy $E_{\rm F}$, the $N=1$ model was fixed to $\varepsilon_{\rm F}$ and to the energy of the bottom of the bonding $pd$--bands at around $-8$ eV. By having to span a wider energy range, the $N=1$ orbitals are somewhat less localized, and consequently have longer--ranged hoppings than the $N=0$ orbitals. A thorough discussion on the relations and trends of the hopping parameters can be found in \cite{kent08} and \cite{Nuss}. In our study we will consider first the $N=0$ and then the $N=1$ model in order to address the following issue of central interest:\\
The most problematic step in recent studies \cite{medici05,kotliar_emery}
was the value of the $d$--$p$ splitting $\varepsilon_d-\varepsilon_p=\Delta_{dp}$. 
While the NMTO downfolding \cite{kent08} yields a value of $\Delta_{dp}=0.45$ eV ($N=0$) or
$\Delta_{dp}=0.96$ eV ($N=1$), it turns out that the many--body treatments, which
include correlation effects fail to reproduce the insulating behavior of the
undoped LSCO. In order to fix this problem $\Delta_{dp}$ was increased ``by hand'' to values of the order of $\Delta_{dp}\approx 3$ eV \cite{kent08, kotliar_emery} or it was chosen as a variable parameter \cite{medici05}. Further, Kent \emph{et al.} \cite{kent08} pointed out that previous justifications of such enhancement of $\Delta_{dp}$ by means of constrained LDA calculations for La$_{2}$CuO$_{4}$ are problematic due to a misleading assumption of the electron count.\\
Our analyis in the subsequent sections will clarify, how these problems are actually related to the assumption of negligible Coulomb interactions between $d$ and $p$ electrons in the many-body calculations. 

\section{Methods: $dp$-models in DMFT+Hartree}
\label{Sec:methods}
We supplement the one-particle extended Emery model from Table \ref{emeryham}
by the following two-particle interactions
\begin{equation}\label{dpHub}
 H_U = U_{dd} \sum_{i} \,  n_{i d \uparrow}n_{i d \downarrow} + U_{pp}
\sum_j \, n_{j p_j \uparrow}n_{j p_j \downarrow} + U_{pd} \sum_{\langle i j \rangle \sigma \sigma'}  n_{i d \sigma}  n_{j p_j \sigma'} \; .
\end{equation}
Here $i$ and $j$  sum over all Cu and O sites, respectively;  $p_j\in\{p_x,p_y\}$
denotes the particular $p$ orbital we include on a given O site $j$
in the Emery model; $\langle i j\rangle$ 
denotes the restriction to nearest neigbors, i.e., a Cu site $i$ and its four surrounding O sites $j$;
and  $n_{i l \sigma}=c^\dagger_{i l \sigma} c^{\phantom{\dagger}}_{i l \sigma}$.
This way the most important (largest) interaction parameters are included: the local Coulomb interactions on the Cu  ($U_{dd}$) and O sites ($U_{pp}$) and
the nearest neighbor  Cu-O interaction $U_{pd}$.

Since part of  $H_U$ is  contained in the NMTO one-particle energies, we need a so-called double counting corrections (DC).  For the $dp$-models, the DC effect is much more important than for $d$-only models where it often corresponds to a simple energy shift that can be ``absorbed'' by the chemical potential. We employ the Anisimov DC formula \cite{AnisimovDC} (also coined as fully localized limit) and extend it straightforwardly to the $dp$--basis:\\
\begin{equation}\label{DC_ani2}
\Delta\varepsilon_{\rm
DC}^{d(p)}={U}_{dd(pp)}\left[n_{d(p)}-\frac{1}{2}\right]+ 4(2) \times {U}_{pd} \, n_{p(d)} \; .
\end{equation}\\
Here, $n_{d(p)}$ are the LDA density for the $d(p)$ orbitals (Cu (O) sites have four O (Cu) neighbors).  So the DC corresponds to a relative shift between the $d$-- and $p$--states, which would be
reversed by a simple mean field treatment of   ${U}_{dd}$ ${U}_{pd}$, and  ${U}_{pd}$.

Let us emphasize that, in contrast to standard LDA+DMFT implementations\cite{explanationLaNiO3},  we treat \emph{all} density-density interactions within the $dp$ basis in  DMFT+Hartree. Specifically,  $U_{dd}$ is treated in DMFT(CTQMC) where we define a local Anderson impurity model from the local Cu $d$ Green function and self energy. The interactions $U_{pp}$ and $U_{pd}$ are treated on the Hartree level, i.e., a self energy contribution for each $d$ and $p$ site 
is calculated as  $U_{pp}$($U_{pd}$) times the  density of the other orbitals involved. Please note  that if $n_{d(p)}$ remained at its  LDA value the DC term would cancel the effect of $U_{pd}$ in DMFT+Hartree exactly. However, due to the
electronic correlations, in particular due to the splitting of the $d$-orbitals,
$n_d$  is reduced and $n_p$ enhanced.

For $U_{pd}$ this treatment is exact in the DMFT limit since non-local interactions reduce to their Hartree contribution, see Ref.~\cite{MuellerHartmann}. In the case of $U_{pp}$, a DMFT solution for the O sites would be in principle required, exactly as for the Cu site. However, since the $p$-orbitals are almost completely filled, electronic correlations are actually weak and the simpler Hartree treatment well reproduces the DMFT solution, as we explicitly show in the Appendix. Our approach incorporates therefore the advantages of the DMFT and  explicitly keeps the $p$--degrees of freedom also in the interaction part of the many body Hamiltonian. The extended basis-set allows us not only to capture  Mott--Hubbard physics, but also that of charge transfer insulators and the concomitant $d$--$p$ interplay. The latter is, in fact, the relevant one for the physics of undoped/underdoped cuprates.

Before turning to the results let us mention that the analytic continuation of the numerical LDA+DMFT results  is more complicated for the  extended $dp$  basis-set. This issue is discussed  in the Appendix as well.   

\section{Results}
\label{Sec:results}
Our DMFT+Hartree results for both 0MTO and 1MTO basis sets can be summarized in two conclusions:
\begin{enumerate}
\item The $d$--$p$ interaction $U_{pd}$, which leads to a \emph{self consistently determined} level shift, drives the system insulating within  DMFT+Hartree and opens the charge transfer gap.\\
\item The critical interactions for this metal-to-insulator transition
 in 0MTO agree with estimates of the interaction strength; for 1NMTO they are at the upper borderline or  somewhat larger than what can be expected.
\end{enumerate}

\begin{figure}%
\includegraphics[width=0.8\columnwidth]{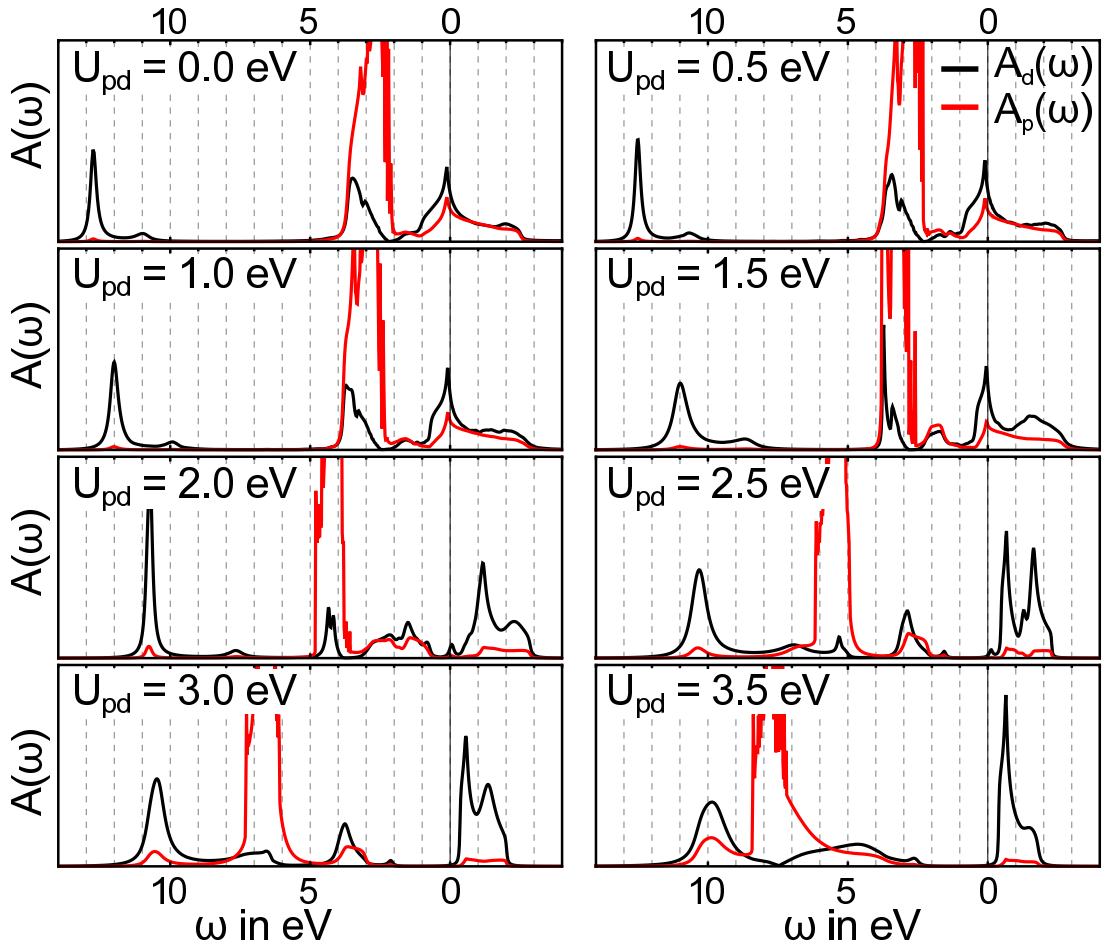}%
\caption{(color online) DMFT spectra for the 0MTO basis with interaction
    parameters $U_{dd}=10$eV, $U_{pp}=5$eV, and varying
    $U_{pd}$.}%
\label{N0MTOgrid}%
\end{figure}

\paragraph{0MTO model} For the interaction parameters of the 0MTO calculation we choose $U_{dd}=10$ eV, $U_{pp}=5$ eV, and take different values of $U_{pd}$ ranging from $U_{pd}=0$ eV to $U_{pd}=3.5$ eV.\\ 
The choice of $U_{dd}$ value is motivated from constraint random phase (cRPA) calculations, which gives $U_{dd}^{\rm cRPA}\approx 9.0\,$eV \cite{hansmann_vaugier}. Such cRPA values are typically smaller than constraint LDA values, which in some cases yields better results in LDA+DMFT calculations. 
The reason for this is the frequency dependence of the $U_{dd}^{\rm cRPA}$ which 
can be actually translated to a Bose factor renormalization of the bandwidth \cite{BF}. To the best of our knowledge, for $U_{pd}$ and $U_{pp}$ no reliable estimates are yet available in the literature, but it is certainly a reasonable assumption that these interactions need to be smaller than $U_{dd}$.
 
The obtained spectral functions are shown in Fig.~\ref{N0MTOgrid}. As in the
bandstructure plots, the color of the spectral function codes the orbital
character. Let us start from the spectrum for $U_{pd}=0$ eV which is plotted in
the top left panel. Although the parameters $U_{dd}=10$ eV and $U_{pp}=5$ eV
are by no means small compared to the bandwidth, we observe a rather
uncorrelated spectral function which resembles the non--interacting DOS
(Fig.~\ref{NMTO_bands}) except for the LHB located below -10eV. The main reason for this is that,
due to the $d$--$p$ hybridization, the filling of each band and, in particular of
the $d$--band, is far from an integer value. Upon increasing the value for
$U_{pd}$ this hybridization decreases, as can be seen in the spectra, since
the charge transfer from $d$--states to $p$--states is now connected with a
potential shift of the respective states of the order of $U_{pd}$.
 Eventually,
for values $U_{pd}\approx \frac{1}{2}U_{pp}$, we observe a rather sudden metal-to-insulator 
transition between $U_{pd}=2$ eV and $U_{pd}=2.5$ eV.

 In
Fig.~\ref{N0MTOgrid} we observe, that for $U_{pd} \leq 2$eV the spectra
show a gap between $d$--states: an ``upper Hubbard band'' with some
$p$--hybridization above the Fermi energy $\varepsilon_{\rm F}$ and a mixed
$d$--$p$ peak around $-2$ eV (for $U_{pd}=2$ eV) or $-3$ eV (for $U_{pd}=2.5$
eV).
The ``lower Hubbard band'' is quite broad and located below $\sim -10$ eV,
whereas most of the $p$--spectral weight is located in a large peak around $-4$
eV ($U_{pd}=2$ eV) and $-5$ eV ($U_{pd}=2.5$ eV). For $U_{pd}=3.5$ eV, the $p$
bands are shifted to such low energies that they  approach the lower Hubbard
bands and start hybridizing with them. Let us emphasize that the formation of the
narrow peak at $E_F$ and the transfer of spectral weight to the lower and upper
Hubbard bands occurring between $2$ and $2.5$eV is a dynamical effect beyond
static mean-field, as witnessed by the strong $\omega$--dependence of $\Sigma$ (see
Fig. \ref{comparesw} below).\\
In summary, DMFT+Hartree yields an insulating state for the original 0MTO parameters \emph{without} the artificial enhancement of the $d$--$p$ splitting hitherto employed in the literature. Instead, we assumed a finite value of $U_{pd}$ which,
\emph{in a self consistent way}, leads to a suppression of $d$--$p$ hybridization driving the metal-to-insulator transition. Yet, we should remark two issues: Since 0MTO was designed to reproduce the cuprate bands only around the Fermi energy \cite{kent08}, it is questionable if the 0MTO really yields a good basis for a study of excitations on an energy scale of some eV above and below the Fermi energy such as the $d$--$p$ interplay. We hence present DMFT+Hartree results of the 1MTO model  in the following.\\

\begin{figure}%
\includegraphics[width=0.8\columnwidth]{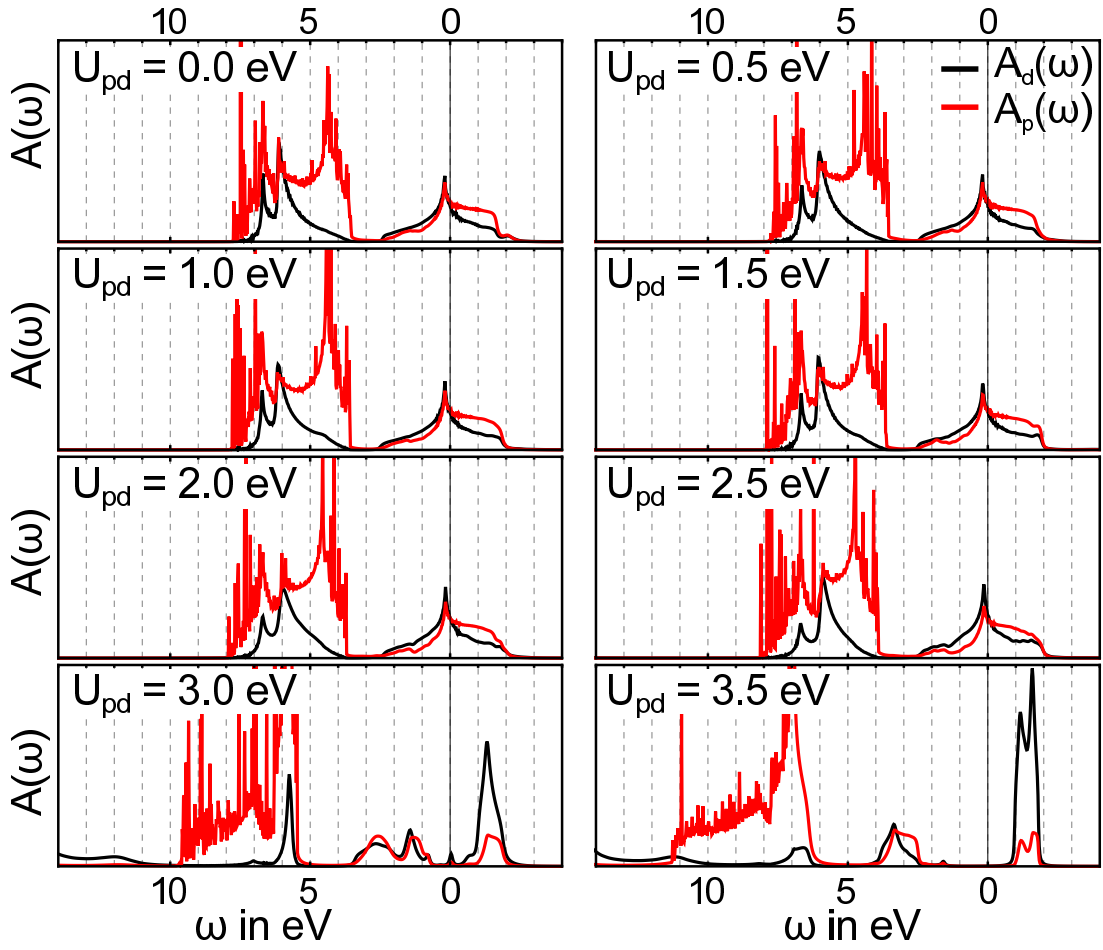}%
\caption{(color online) DMFT spectra for the 1MTO basis with interaction
    parameters $U_{dd}=13$eV, $U_{pp}=5$eV, and varying
    $U_{pd}$. We have not broadened the $\delta$--peaks with an additional
    Lorentzian broadening. The only broadening is due to the intrinsic finite
    lifetime from ${\rm Im}\Sigma$. Please note that there is spectral weight down
to $-20$eV in the lower two panels which we cut off.}%
\label{N1MTOgrid}%
\end{figure}

\paragraph{1MTO model} 
As mentioned above, the N=1 orbitals are less localized, and consequently have somewhat longer--ranged hoppings than the N=0 orbitals in order to reproduce the band structure on a larger energy range. Hence, also the corresponding 1MTO values of the local $U_{dd}$ and $U_{pp}$ interaction parameters should be reduced. Concurrently, we find that the 1MTO yields much more \emph{metallic solutions}. The system remains metallic even for the same interaction parameters as we used for the N=0 model in Fig.~\ref{N0MTOgrid}; to obtain an insulating ground state one needs a larger  $U_{dd}$ or  $U_{pd}$.

In Fig.~\ref{N1MTOgrid} we show the DMFT+Hartree spectral functions for the N=1 model at  $U_{dd}=13$ eV, $U_{pp}=7$ eV, and varying $U_{pd}$. Increasing $U_{pd}$ beyond   $3$ eV opens   a gap  in the spectral function However,  $U_{dd}=13$ eV is at the upper edge or larger than the physically reasonable parameter range.

\begin{figure}[t]%
\includegraphics[width=\columnwidth]{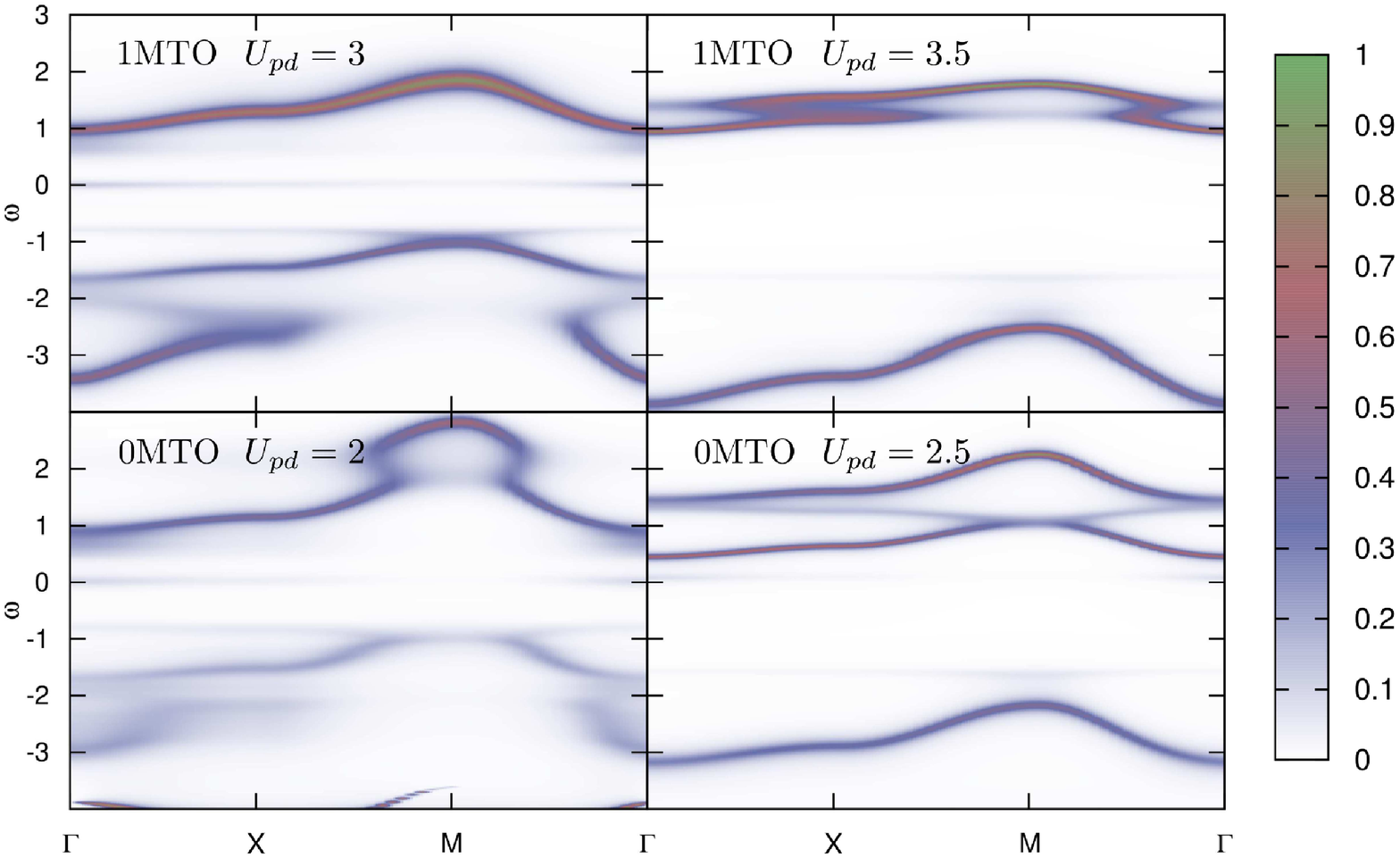}%
\caption{(color online) $\vec{k}$-resolved spectral function of 0MTO in the lower
    and the 1MTO in the upper row. The two left panels are before
    the charge-transfer metal-insulator transition (i.e., for $U_{dd}=10,\,U_{pp}=5,\,U_{pd}=2$ eV and $U
_{dd}=13,\,U_{pp}=7,\,U_{pd}=3$ eV in the lower and upper row, respectively).
The two right columns are after the metal-insulator transition (obtained by increasing $U_{pd}$-parameter by 0.5 eV; keeping the other parameters fixed). For both
    models the mixed $pd$-state becomes suddenly coherent as soon as we enter 
the insulating state.}
\label{kresolved}%
\end{figure}

\paragraph{Nature of the  mixed $d$--$p$ peak}  As shown in Figs. \ref{N0MTOgrid}, \ref{N1MTOgrid} a peak of mixed $d$- and $p$-orbital character develops in the  $\vec{k}$-integrated spectral functions around $-3$ eV (for 0MTO at $U_{pd}=2.5$ eV and 1MTO at $U_{pd}=3.5$ eV).  This peak can be interpreted\cite{note_ZR} as a  Zhang-Rice state in agreement with  previous LDA+DMFT studies \cite{Yin2008,Weber2010,wang11}.

 In order to see whether or not this spectral weight is also associated to a coherent excitation we calculate $A(\vec{k},\omega)$ just before and after the metal-insulator transition, i.e., $U_{pd}=2-2.5$ eV for 0MTO and $U_{pd}=3-3.5$ eV for 1MTO. The results are shown in Fig. \ref{kresolved}, and the corresponding local $d$-self-energies are shown in Fig. \ref{comparesw}.
Interestingly, the coherence of the spectral feature around $-3$ eV changes dramatically across the metal-insulator transition for both 0MTO (the two lower panels of Fig. \ref{kresolved}) and 1MTO (the two upper panels).
In the two left panels of Fig. \ref{kresolved}, i.e., before the metal-insulator transition, the dispersive band between $-3$ and $-1$ eV is very incoherent whereas in the two right panels, i.e., just after the metal-insulating transition, a very well defined quasiparticle excitation emerges. 
The double-occupancies of the $d$-orbital and the spectral weight at the Fermi level as a function of $U_{pd}$ shows the following:
The values of $U_{pd}$ for which the double-occupancies (not-shown) get suppressed are those for which most of the spectral weight gets shifted away from the Fermi level. In this strongly-correlated metallic solution an additional peak of mixed $dp$ character is formed when a residual finite fraction of itinerant electrons is still present. This spectral feature, which can be interpreted as the emergent Zhang-Rice excitation, becomes more coherent only when the spectral weight at the  Fermi level is completely depleted. This depletion, i.e., the opening of the gap, occurs in a similar way as in the Mott-Hubbard transition of the Hubbard model: a pole in the real part of the self-energy develops, see the blue (dashed) lines
in Fig. \ref{comparesw}. At the same time, the imaginary part of the self energy develops a peak.
 Since the gap is between $d$-states above and $p$-states below the Fermi level,
we have however a charge-transfer insulator\cite{ZSA} with a  Zhang-Rice  singlet.


\section{Conclusions}
\label{Sec:conclusion}
\begin{figure}%
\includegraphics[width=\columnwidth]{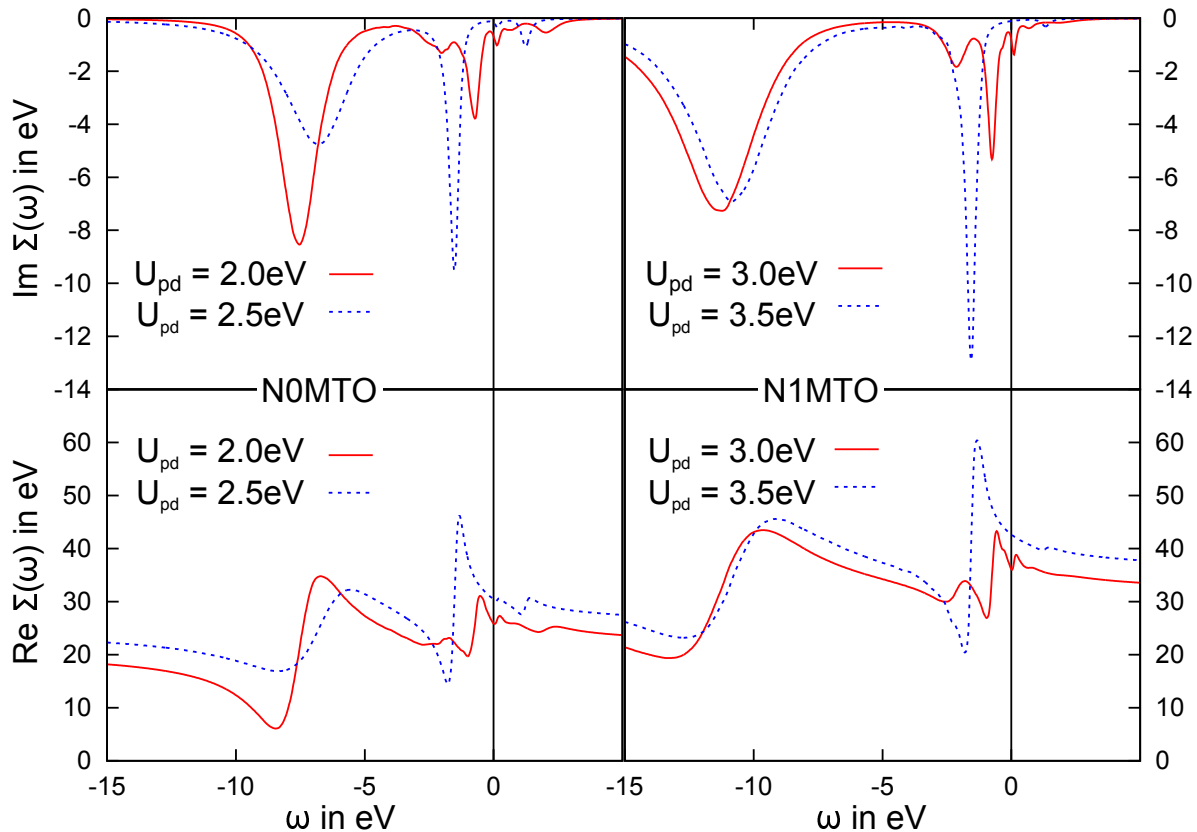}%
\caption{(color online) The imaginary  (upper panels) and real part (lower panels) of the local DMFT $d$-self-energy on the real axis for the four cases plotted in Fig. \ref{kresolved}.}%
\label{comparesw}%
\end{figure}

Presently, within the state-of-the art LDA+DMFT calculations, the splitting between the $d$- and $p$-states in
cuprates \cite{kent08} and other transition metal oxides is adjusted for obtaining agreement with experiment, see e.g. Ref.~\cite{Millis13},
or new double counting corrections are introduced, see Ref.~\cite{Gabi13}. This is quite unsatisfactory since it basically destroys the often claimed {\sl ab initio} character of the calculation. 
The aim of our work  was hence (i)  to understand the physical origin behind such adjustments, and (ii) to perform calculations based on the original parameters from NMTO downfolding \emph{without} artificially changing them.  In particular, if the orbital overlap is large, the non-local oxygen-copper interaction $U_{pd}$ can be strong. We have shown that the inclusion of this interatomic interaction is indeed extremely important, because it allows for the description of a \emph{self consistently determined $d$-$p$ level splitting}. This mechanism can explain why  previous LDA+DMFT calculations that only included the $d$-$d$ interaction yielded -in some cases- poorer results than corresponding studies in smaller $d$-only basis-set. The  artificial $p$-$d$ shifts that were employed in the literature  actually mimic the effect of $U_{pd}$.  The DMFT+Hartree approach with the explicit inclusion of $U_{pd}$, adoped in this work, is much more satisfactory from both a practical and a conceptual point of view. The DMFT+Hartree treatment can be based entirely on interaction parameters determined {\sl ab-initio} for the downfolded basis-set chosen, e.g., by means of cRPA\cite{Ferdy2004} or its recently developed locally unscreened version\cite{Nomura2012}.    

For the specific case of undoped cuprates, the two downfolded models considered, 0MTO and 1MTO, qualitatively yield the same physics, see Figs.\ 2-4. Quantitatively, the results for the 1MTO model require somewhat larger interaction parameters than 0MTO to open an insulating gap. 
For 0MTO we have a charge transfer insulator for plausible
interaction parameters, whereas for 1MTO these are  at the borderline
or even somewhat larger than expected \cite{hansmann_vaugier}.  
Let us note in this respect that non-local interactions, beyond the realm of our DMFT+Hartree treatment,
also play an important role for cuprates and
will further stabilize the insulating solution.
Also in the paramagnetic phase these non-local correlations
will reduce the critical interactions for the metal-to-insulator transition\cite{Uc-2D,Comanac}.

Let us emphasize  that the impact of our analysis goes well beyond the particular, though significant case of the insulating phase in undoped cuprates. In fact, for other transition metal oxides with a more three dimensional crystal and electronic structure, non-local correlations beyond DMFT
are less important, except for low temperatures\cite{DGAresul,Gull3D}.  Hence, one can expect a DMFT+Hartree calculation including {\sl ab-initio} the $U_{pd}$ interaction to be sufficient for capturing the physics of downfolded $d$-$p$ models which are nowadays fixed by hand\cite{Millis13}. In practice, similarly as for the Emery model, the $U_{pd}$ interactions will turn the physics of the $dp$-model towards that of the $d$-only model, because the number of $d$ electrons is reduced towards an integer value, entailing stronger electronic correlations. Our paper hence paves the way 
for better treatment of oxygen $p$-orbitals in LDA+DMFT calculations.

\section*{Acknowledgements}
We acknowledge very helpful discussions with O. K. Andersen and T. Saha-Dasgupta as well as L. Vaugier, P. Seth and S. Biermann. We also would like to thank F. F. Assaad for
providing us with the maxent code.
A.T. benefited from financial support by the  Austrian Science Fund (FWF)
through  research unit FOR 1346 (FWF project ID I597); K.H. from the FWF
SFB ViCoM F41.
G.S. acknowledges support by the Deutsche Forschungsgemeinschaft (FOR 1162).
Calculations have been done on the Vienna Scientific Cluster.

\section*{Appendix}

\paragraph{Analytic Continuation}

To obtain high quality spectra with as little ambiguity as possible we employ the continuation of the self-energy scheme as described in \cite{wang2009}. Compared to the more often used continuation of the Green function, the continuation of the self-energy has the advantage of leading to guaranteed physical self-energies, since the condition of positivity of the imaginary part of the self-energy is incorporated in the maximum entropy method (maxent). Furthermore, compared to the maxent for the Green functions, this scheme has the advantage of avoiding an unnecessary maxent-step for the $p$-bands, whose self-energy is known and constant.
This way maxent is applied only to the part beyond Hartree of the correlated orbitals, whereas all other quantities, e.g. $H_{\vec{k}}$ and the Hartree self-energy $\Sigma_\text{H}$ can be treated exactly on the real axis. \\
In the practical implementation, we measure the Greens function from a converged run directly in Matsubara frequencies $i\omega_n$ to avoid binning errors as would be the case when measuring the Greens function in imaginary time $\tau$. Then we calculate the self-energy $\Sigma(i\omega_n)$ using the Dyson equation $\Sigma(i \omega_n) = {\cal G}_0^{-1}(i\omega_n)-G^{-1}(i\omega_n)$, assuming that the bath Greens function ${\cal G}_0^{-1}$ does not contain statistical errors. To calculate the error for the self-energy we employ the bootstrap method on 300 independent Greens function measurements.\\
The self-energy has a different asymptotic behavior from that of the Greens function. The latter is however needed for maxent-like methods. Therefore a manipulation of the input is needed. To this end one performs a high-frequency expansion of the CTQMC self-energy which for the single band Hubbard model reads:
\begin{align}
    \label{selfenergyhighw}
    \Sigma_\sigma(i\omega_n)=&U_{dd}\langle n_{-\sigma}\rangle \\ &+ U_{dd}^{2}\langle
    n_{-\sigma}\rangle (1-\langle n_{-\sigma}\rangle) (i\omega_n)^{-1} +
    \mathcal{O} \left( ( i\omega_n)^{-2} \right). \nonumber
\end{align}
Following the argument above and with the knowledge that the Greens function
has a $1/(i\omega_n)$ asymptotic behavior, we subtract the Hartree term for the
$d$ band, $U_{dd}\langle n_{-\sigma}\rangle$ and divide the remaining part of
the self-energy by the prefactor of the $1/(i\omega_n)$ term. This new object,
$\widetilde{\Sigma}(i\omega_n)$, can then be analytically continued to the real
axis by means of the stochastic maxent method \cite{beach04,abendschein06}.

In this respect, by inspecting the correlation matrix of the self-energy we checked that different Matsubara frequencies are uncorrelated, otherwise a transformation to a non-correlated basis would have been necessary. 
We finally calculated the spectral function, $A(k,\omega)$, from $\widetilde{\Sigma}(\omega)$ and all other terms, namely the double counting correction $DC$, the $\vec{k}$-dependent Hamiltonian $H_{\vec{k}}$, the chemical potential $\mu$ and the Hartree self-energy of $\Sigma_\text{H}$:
\begin{align}
    A_{\sigma}(\vec{k},\omega)  = -\frac{1}{\pi} \text{Im} & \Big\{  \omega + \mu - DC - H_{\vec{k}} -
	\Sigma_\text{H}  \nonumber\\
     &  - \widetilde{\Sigma}(\omega)U_{dd}^{2}\langle
	n_{-\sigma}\rangle (1-\langle n_{-\sigma}\rangle) \Big\}^{-1}
    \label{spectralfunction}
\end{align}
To test the validity of our approach we also compared data of an ED-calculation with data obtained by our CTQMC implementation, which resulted in very good agreement.

\paragraph{DMFT(CTQMC) for the p-electrons}

\begin{figure}
   \captionsetup[subfigure]{labelformat=empty}
\centering
\subfloat[]{\includegraphics[width=0.5\textwidth]{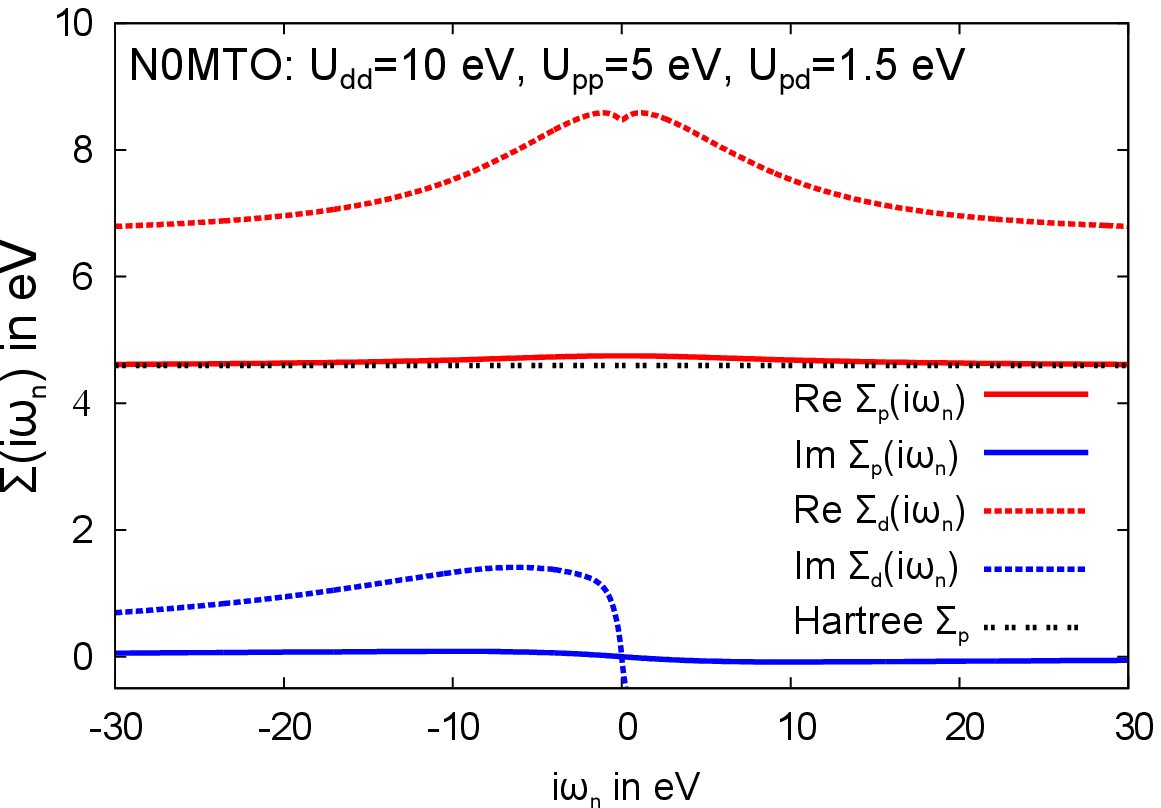}}
\hfill
\subfloat[]{\includegraphics[width=0.5\textwidth]{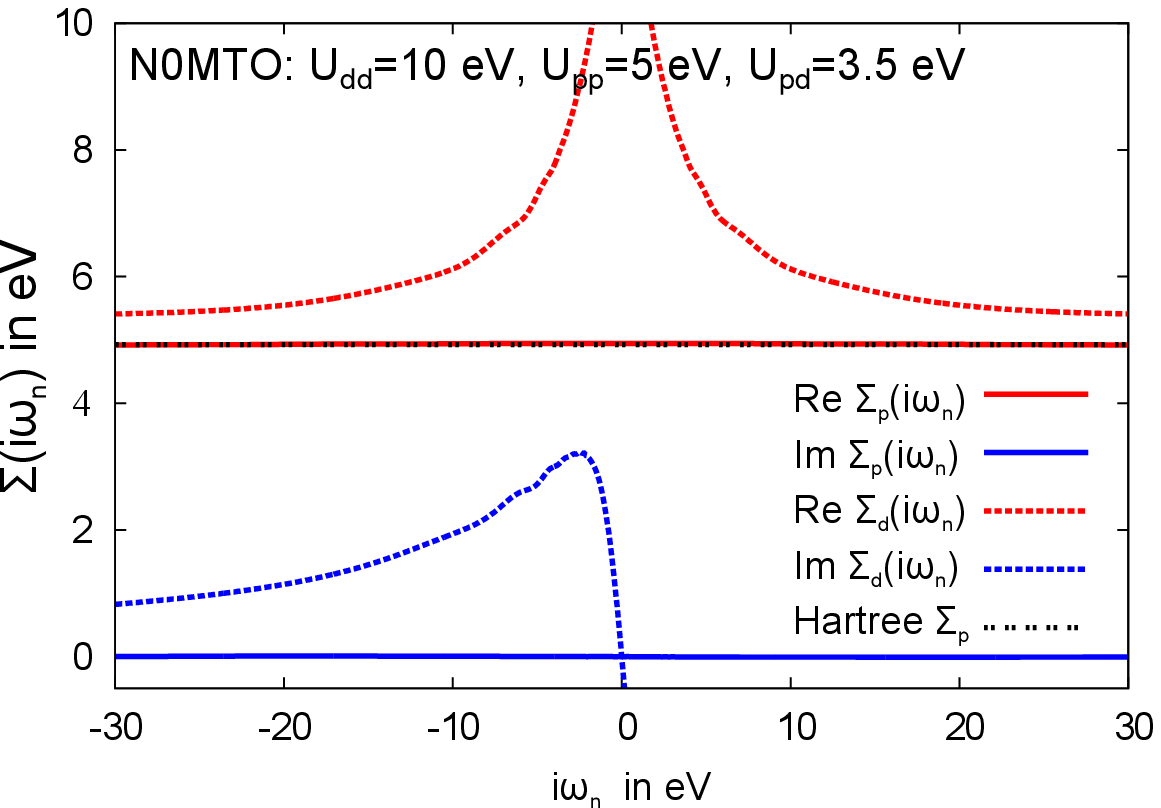}}
\caption{(color online) Local DMFT self-energy of the $p$ orbitals compared to the corresponding Hartree calculation
for the 0MTO Emery model with $U_{pd}=1.5(3.5)\,$eV, $U_{pp}=5\,$eV,
$U_{dd}=10\,$eV 
in the left (right) panel. For both cases the Hartree self-energy of the
$p$ orbitals is almost identical to the one which was calculated with the CTQMC.}%
\label{compareHartreeDMFT}%
\end{figure}

Here, we  validate that a Hartree treatment of $U_{pp}$ is sufficient. To this end we performed  for some test  calculations where
both $U_{dd}$ and $U_{pd}$ are treated in DMFT, and $U_{pd}$ in Hartree, which actually is the correct (full) DMFT treatment of the Emery model.
Fig.\ \ref{compareHartreeDMFT} shows the obtained DMFT self energy, which is basically the same as the
Hartree self energy.  For  $U_{pd}=1.5\,$eV,  very minor deviations are discernible at low frequencies,
but the Hartree self energy is still an accurate approximation of  the DMFT self energy. Therefore, we conclude that a Hartree treament of $U_{pp}$ is absolutely sufficient.


\begin{thebibliography}{99}
\bibitem{note_ZR} Note that DMFT allows for the description of a singlet state between the $d$-site and the bath electrons of the associated Anderson Impurity model. In the specific case considered, as it is typically done, the  dynamical Weiss field of DMFT effectively includes also the $p-$degrees of freedom, which are therefore not explicitly separated. The strong mixing among $d$ and $p$ orbitals observed at $-3$eV, however, supports the intepretation of the corresponding spectral feature as Zhang-Rice state.
\end{thebibliography}
\end{document}